\documentclass[preprint,showpacs,aps]{revtex4}

\usepackage{graphicx}
\usepackage{dcolumn}
\usepackage{bm}

\begin{document}
\title {Irreversibility line of the MgB$_{2}$ superconductor}
\author {Y.Z. Zhang$^{1,2}$, R. Deltour$^2$, H.H. Wen$^1$, J.F. de Marneffe$^2$, H. Chen$^{1}$, S.F. Wang$^{1}$, S.Y. Dai$^{1}$, Y.L. Zhou$^{1}$, Q. Li$^{3}$, A.G.M. Jansen$^{4}$, P. Wyder$^{4}$, and Z.X. Zhao$^1$}
\affiliation{$^1$National Laboratory for superconductivity \& Laboratory of Optical Physics, Institute of Physics \& Centre for Condensed Matter Physics, Chinese Academy of Sciences, P. O. Box 603, Beijing 100080, China\\ $^2$Universit\'{e} Libre de Bruxelles, Physique
des Solides, CP-233, B-1050, Brussels, Belgium 
\\$^3$Department of Physics, Pennsylvania State University, University 
Park, PA 16802, Pennsylvania, USA\\$^4$Grenoble High Magnetic Field Laboratory, Max-Planck-Institut f$\ddot{u}$r 
Festk$\ddot{o}$perforschung and Centre National de la Recherche Scientifique, 25, 
Avenue des Martyrs, B. P. 166, F-38042 Grenoble Cedex 9, France}
\begin{abstract}
Magneto-resistivity of a $c$-axis oriented MgB$_{2}$ thin film was studied in perpendicular and parallel magnetic fields up to $\sim $23 T with temperatures down to 0.38 K. Resistive critical magnetic fields were determined. Large separations between irreversibility lines and upper critical magnetic fields are observed. An effective quantum parameter as a function of temperature is proposed to explain the separations under the theoretical framework of quantum fluctuations, in good agreement with our experimental and previously published experimental data.
\end{abstract}

\maketitle
The discovery of superconductivity in MgB$_{2}$ \cite {Nagamatsu} has stimulated enormous activity in the study of this material as it is certainly promising for future applications. One of the interesting phenomena is that the irreversibility line $H_{irr}(T\to 0)$ of MgB$_2$ does not show an upward curvature as observed in the studies of high $T_c$ superconductors (HTSCs) \cite {Cohen,Ando}. $H_{irr}(T)$ is defined as the critical field beyond which persistent supercurrent can not be sustained and energy dissipation is present in the superconductor. For a HTSC, thermal fluctuations are much greater, which lead to easier flux line motion near the upper critical magnetic field $H_{c2}$, thus resulting in a large separation between $H_{irr}(T)$ and $H_{c2}(T)$ at sufficiently high temperatures. However, at low temperatures for HTSCs \cite {Cohen,Ando}, a steep rise of $H_{irr}(T)$ leading to a drastic reduction of the separation from $H_{c2}(T)$ is widely observed. For MgB$_2$, it is known that the separation between the two fields \cite {Eom,Kim,Wen,Wen2,Patnaik,Buzea}
increases with decreasing temperature. However, $H_{irr}(T)$ measurements at  very low temperatures ($T<1.6$ K) are still missing. In this paper, the results of a $c$-axis oriented MgB$_{2}$ thin film \cite {Wang} are reported in magnetic fields up to 23 T and temperatures down to 0  .38 K. Experimental data show that the different values of the resistive determined $H_{irr}(T)$ and $H_{c2}(T)$ are nearly constant for $T < 1.5$ K. Based on these data and previously published experimental data \cite {Wen},  the temperature dependence of an effective quantum parameter is proposed to explain the $H_{irr}(T)$ anomaly within the theoretical framework of quantum fluctuations \cite {Blatter}. 

A boron thin film was first deposited on (0001) oriented Al$_{2}$O$_{3}$ single crystal substrate and then followed by an \textit{ex situ} annealing in Mg vapor at $900^{\circ}$ C for one hour \cite {Wang}. Such a film of thickness around $\sim $500 nm can reach a superconducting transition temperature of 39 K. 
This film is highly $c$-axis oriented with the full width at half maximum of the (002) rocking curve being $\sim 1.8^{\circ}$. \textit{AC}-susceptibility measurements show that the film studied in this experiment has a transition temperature $T_{c} \approx 33$ K. For magneto-resistivity measurements, a bridge geometry was patterned on the film with dimensions of 0.2 $ \times $ 2 mm$^{2}$. A four-probe \textit{AC} technique was used with a current modulation at 91.05 Hz (magnitude 4.0 $\mu$A). Perpendicular and parallel magnetic fields to the thin film plane were sequentially applied with the current always perpendicular to the field directions. Magnetoresistivity data were collected with a field sweeping rate 3.0 x10$^{-2}$ T/s at fixed temperatures. 

\begin{figure}
\includegraphics
[width=0.88\columnwidth]
{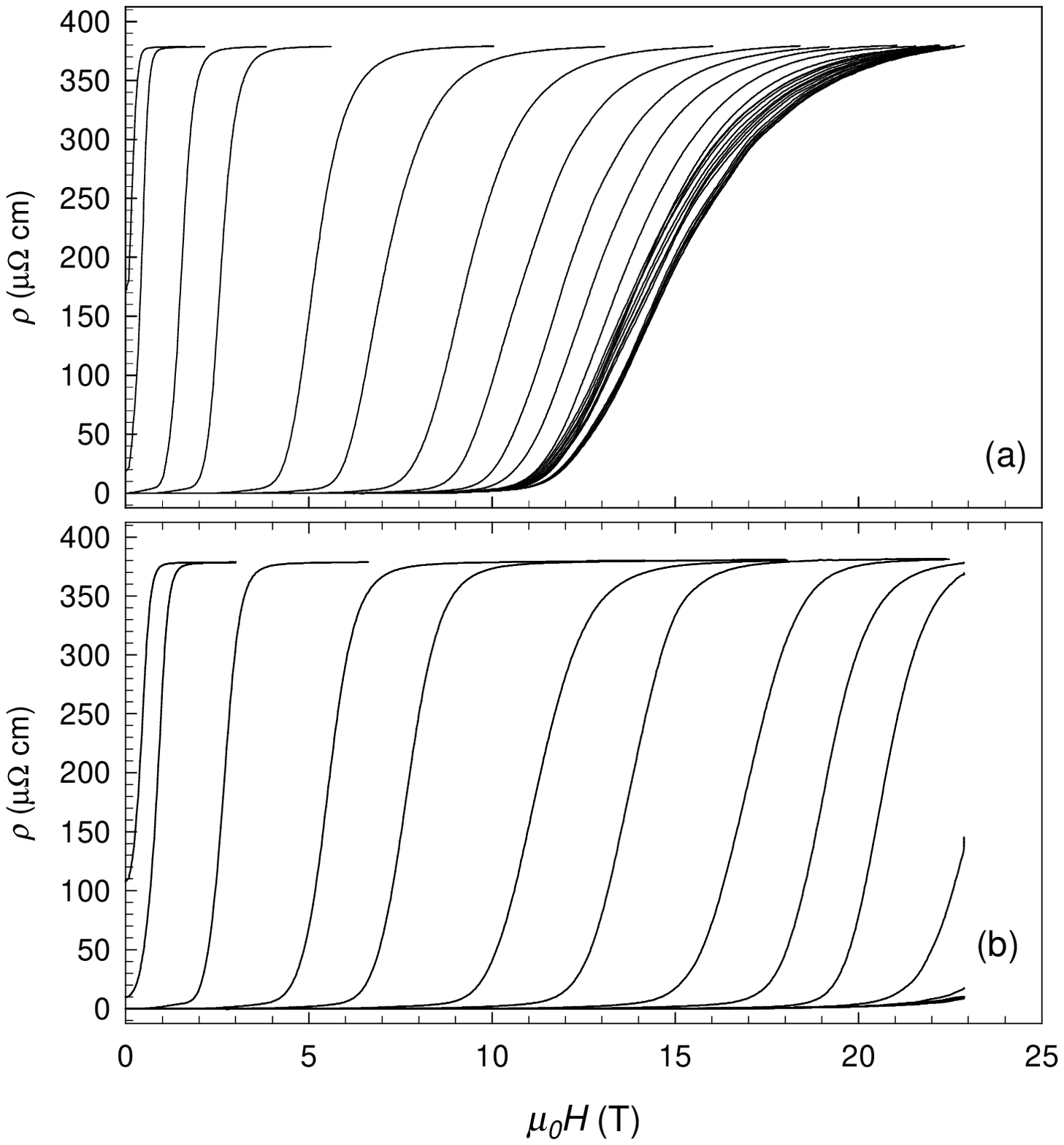}\caption{\label{f1} Resistivity isotherms of the thin film in (a) perpendicular fields (from right to left: $T =$ 0.38, 0.43, 0.61, 0.85, 1.17, 1.49, 1.70, 
1.94, 2.16, 2.52, 3.26, 3.41, 3.78, 4.4, 5.6, 7.4, 9.0, 11.5, 14.3, 17.5, 
22.9, 26.8, 30.1, 30.9, 32.6, 33.8 K), and (b) parallel fields (from right 
to left: $T$ = 0.41, 0.82, 2.06, 3.93, 4.6, 7.4, 9.1, 11.2, 14.7, 
19.2, 22.8, 26.3, 28.3, 30.5, 32.4, 33.4 K).}
\end{figure}

\begin{figure}
\includegraphics
[width=0.88\columnwidth]
{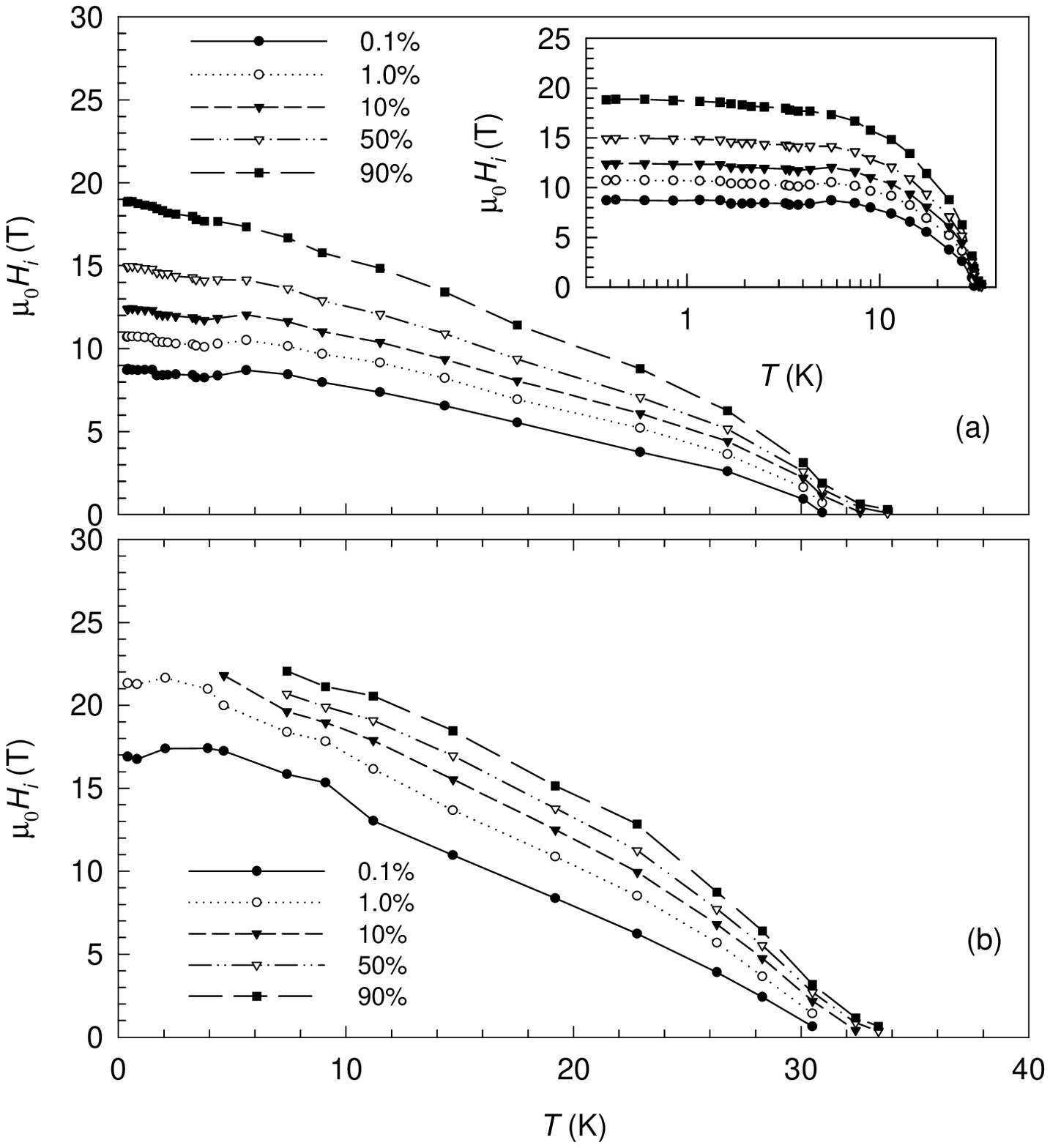}\caption{\label{f2} Resistive critical fields determined with resistive criterion: 
0.1\%, 1.0\%, 10\%, 50\%, 90\% in (a) perpendicular fields and (b) parallel 
fields. Inset of (a), the corresponding data in a log-linear scale.}
\end{figure}

Figures~\ref{f1}(a) and ~\ref{f1}(b) show the magnetoresistivity curves in perpendicular and parallel fields, respectively. Using a constant resistivity value $\rho_{N}$ in the normal state, we determine transition fields 
$H_{i}(T)$ for different fractions $i = \rho(T,H_{i})/\rho_{N}$ of the normal state resistivity. 
Fig.~\ref{f2}(a) and ~\ref{f2}(b) represent the experimental data $H_{i}^{ 
\perp}(T)$ and $H_{i}^{\parallel}(T)$ for the perpendicular and parallel fields, respectively. Here, we simply define 
$H_{c2}(T) = H_{90\%}(T)$,  
$H_{irr}(T) = H_{0.1\%}(T)$ (a small dissipation criterion), and characterize the magnetic field difference by the normalized parameter $\delta h=[H_{c2} -H_{irr}]/H_{irr}$. 
%
%
%
%
%
%

The inset of Fig.~\ref{f2}(a) shows $H_{i}^{\perp}(T)$ curves in a 
log-linear scale. It is remarkable that each $H_{i}^{\perp}(T)$ curve shows an approximately constant value for temperatures below 1.5 K, which corresponds to the lowest temperature curves in Fig.~\ref{f1}(a) where all the curves roughly fall onto the same line for $T < 1.5$ K. By extrapolating to $T = 0$ K, we obtain $\mu _{0} H_{irr}^{\perp}(0) = 8.7 \pm 0.5$ T, $\mu_0H^{\perp}_{c2} = \Phi_0/2\pi \xi_{ab}^2(0)=18.9 \pm  1.0$ T with the in-plane coherence length $\xi_{ab}(0) \approx 4.2$ nm. Note that $\delta h(0)= [H_{c2}^{\perp}(0)- H_{irr}^{\perp}(0)]/ H_{irr}^{\perp}(0) \approx 1.2$ derived here is very close to those of previous reports, but the $\rho_N$ is about 2 orders larger than that of MgB$_2$ bulk materials ($T_c\approx 39$ K) \cite {Eom,Kim,Wen,Wen2,Patnaik,Buzea}.

From Fig.~\ref{f2}(b), by extrapolation, we obtain $\mu _{0} H_{irr}^{\parallel}(0) = 17.5 \pm 1.0$ T.  Note the $\mu_0H^{\parallel}_{c2}(0)$ is extrapolated from the data above 4.6 K due to the available 
data within the magnetic field limit. The value for $\mu _{0} H_{c2}^{\parallel}(0)= \Phi_{0} /2\pi \xi _{ab}(0) \xi _{c}(0)$ is then $27.5 \pm  2.5$ T with the out-plane 
coherence length $\xi _{c}(0) \approx 2.8$ nm. A rough estimate for the anisotropy factor is $\gamma  = H_{c2}^{\parallel}(0)/H_{c2}^{\perp}(0) \approx 1.5$. 

According to Blatter and Ivlev \cite {Blatter}, the vortex motion may not only 
be related to thermal effects, but also to quantum effects \cite {Brandt,Ivlev,Smith,Blatter1}. 
The vortex displacement amplitude relation for superconductors under the influence of both thermal and quantum fluctuations is given by 
%
%
\begin{equation}
\frac{{ < u^{2} >} }{{a_{0}^{2}} } \approx \left( {\frac{{G}}{{\beta _{th} 
}}} \right)^{1/2}\frac{{\sqrt {b}} }{{1 - t - b}}\left[ {t + q\sqrt {b} 
\left( {1 - \frac{{b}}{{1 - t}}} \right)} \right],
\label{eq1}
\end{equation}
%
%
where $a_{0}$ is the vortex lattice constant, $<u^{2}>$ 
the mean square displacement, $b =B / H_{c2}(0)$, $t =T /T_c$, $q$ the quantum parameter characterizing the quantum fluctuations, $\beta_{th}\approx  2.5$ the numerical parameter, and $G$ the Ginzburg number characterizing the thermal fluctuations. Using this relation, we have two simple approaches for two limits. First, supposing $q$ is large enough (strong quantum fluctuations) for $t\to 0$, we have $< u^{2} > / a_{0}^{2} \approx qb({G/\beta _{th}}  )^{1/2}$, thus the vortex displacement amplitude being only related to the applied magnetic 
field in absence of thermally induced hopping. Second, supposing $q = 0$ 
(no quantum effects), we find $ < u^{2} > /a_{0}^{2} \to 0$ for $t\to 
0$. Note that the term $ < u^{2} > /a_{0}^{2}$ can be used to describe the 
Lindemann criterion (with a value between $0.1\sim 0.3$) for the vortex 
lattice melting. Therefore, we can conclude that in this limit there 
is no vortex lattice melting transition at $T = 0$ even by increasing 
the magnetic field to $H_{c2}(0)$. This second limit, $q = 0$, implies that the vortex melting line $H_{m}(T\to 0)$ approaches $H_{c2}(0)$, and this should also be the case for $H_{i}(T\to 0)$s 
[$H_{i}(T\to 0)\ge H_{m}(T\to 0)$]. In 
the first limit, one expects a weak temperature dependence of  
$H_{i}(T\to 0)$ in accordance with our result. In the 
second limit, $H_{i}(T\to 0)$ should approach 
$H_{c2}(T\to 0)$, rather different from the large separation 
of $H_{irr}$ and $H_{c2}$ we observed. 

Applying the Lindemann criterion $ < u^{2} > /a_{0}^{2} = c_{L}^{2}$ to Eq. 
(\ref{eq1}), where $0.1\le c_L\le 0.3$, Blatter 
and Ivlev \cite {Blatter} obtained the approximate solution for the melting line
%
%
\begin{equation}
H_{m} \approx 4\theta ^{2}H_{c2}(0)/ (1 + \sqrt {1 + 
4S\theta /t})^{2},
\label{eq2}
\end{equation}
%
%
with the temperature variable $\theta  =  c_{L} ^{2} (  {\beta _{th}  /G } 
) ^{1/2 }( { 1/t - 1 }  )$, and the suppression parameter $S = q 
+  c_{L} ^{2}  ( { \beta {th}  / G} )^{1/2 }$. From the theory:
%
%
\begin{equation}
q  = 2\tau_rcQ\sqrt{\beta_{th}/G}/\pi^3\lambda   \approx 2\nu \sqrt {\beta_{th}/G} /\pi^3 K_F \xi,
\label{eq3}
\end{equation}
%
%
where $Q=e^2\rho_N/\hbar d$ is the quantum sheet resistance, $\rho_N= m/e^2n\tau_r$, $d$ the layer spacing, $m$ the effective mass, $\tau_r$ the relaxation time, $n$ the carrier density,  $\lambda$ the penetration depth, $\nu$ the free parameter, and $K_{F}$ the Fermi wavenumber nearly constant at low temperature. 

Taking $H_{irr}\to H_m$ \cite {Blatter1,Suenaga,Schmidt,Blatter2,Houghton}, from Eqs. (\ref {eq2}) and (\ref {eq3}), one can find 
%
%
\begin{equation}
\delta h (0)\approx  q(0)\sqrt{G/\beta_{th}}/c^2_L=2c\tau_rQ/\pi^3\lambda(0)c^2_L
\label{eq4}
\end{equation}
%
%
that means $\delta h(0) \propto \tau_rQ \propto \tau_r \rho_N \propto m/n$. Note that no great $\delta h(0)$ change has been observed for two orders of the magnitude change of the MgB$_2$ resistivity, leading to the conclusion that the products of  $\tau_r\rho_N$ and $mn^{-1}$ of MgB$_2$ are roughly constant. 
Note that, besides the strong difference: $n\sim 10^{21}$ cm$^{-3}$ for HTSCs and $n\sim 10^{23}$ cm$^{-3}$ for MgB$_2$ \cite {Buzea}, steep upward curvatures of $H_{irr}(T)$s of HTSCs were observed at low temperatures \cite {Cohen,Ando}, resulting in relatively small $\delta h (0)$s as compared to that of MgB$_2$. These $\delta h(0)$ and $n$ features lead to a conclusion that the $\tau_r$ and the $m$ of a HTSC are much smaller than that of MgB$_2$. These conclusions rely on the assumption of a single relaxation time $\tau_r$, independent of the vortex state; failures of this assumption shall modify the $\delta h(0)$ relation by a numerical factor accounting for the characteristics.

\begin{figure}
\includegraphics
[width=0.88\columnwidth]
{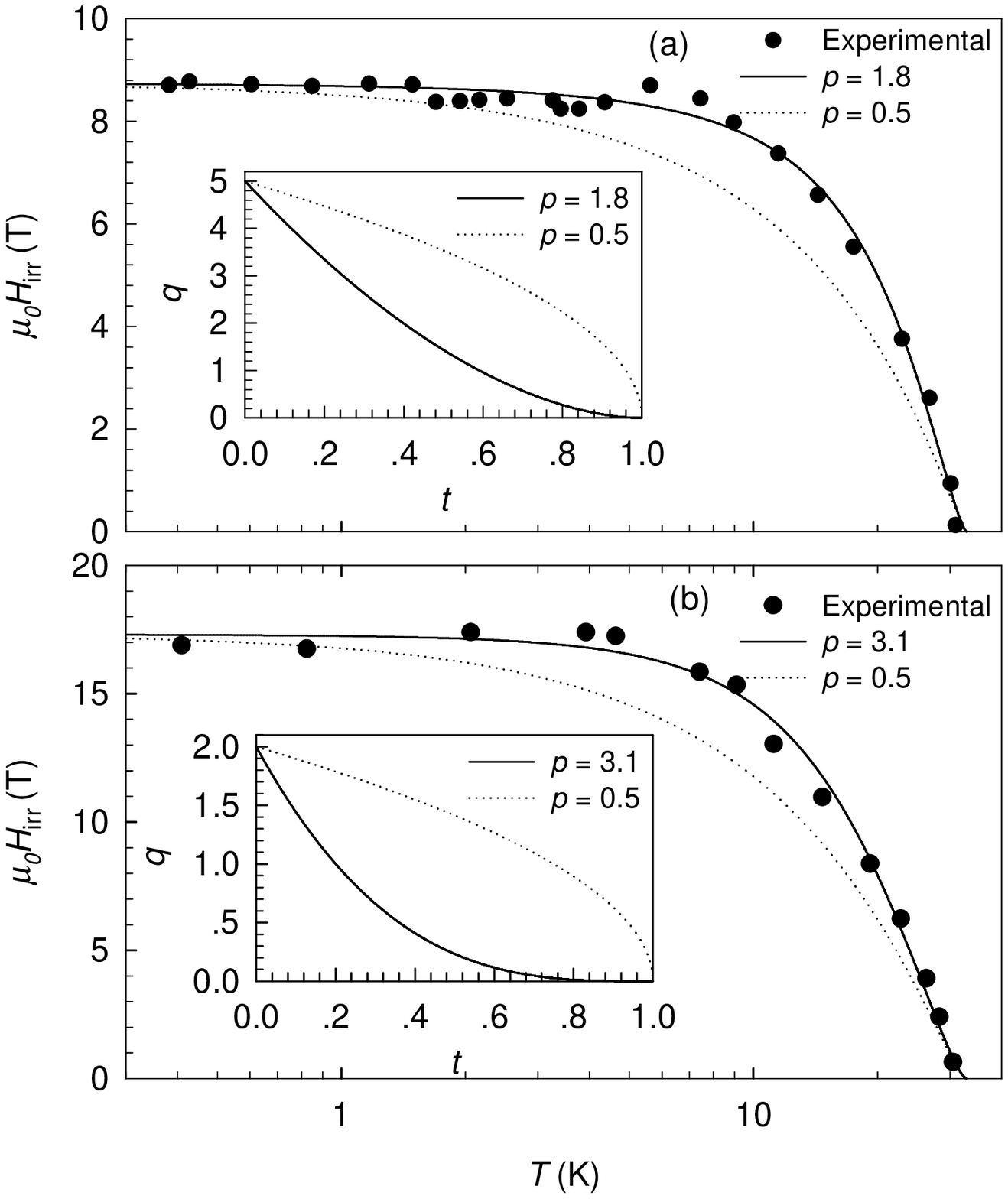}\caption{\label{f3} Experimental data of $H_{irr}^{ \perp}(T)$ in the perpendicular field direction (a), and $H_{irr}^{\parallel}
( {T} )$  in the parallel field direction (b), with theoretical fits: dotted line for the fit with $p=1/2$, solid line for the best fit with the free parameter $p$.}
\end{figure}

Using a temperature dependence of the coherence length, $\xi \approx \xi ({0}) / ({1 - t}) ^{1/2}$, we have:   
$q(t) \approx 2\nu \sqrt {\beta_{th}/G} /\pi^3 K_F \xi \approx q_0( {1 - t} ) ^{1/2}$
with $q_0$ the quantum parameter at 0 K. Hence, we have two free parameters: $q_0$, and $c_f = c_{L}^{2} ( {\beta_{th} /G} )^{1/2}$. With $T_{c} = 33$ K, $H_{c2}^{ \perp} (0) = 18.9$ T and $H_{c2}^{\parallel} (0)= 27.5$ T, we present in Figs.~\ref{f3}(a) and ~\ref{f3}(b) the fitting lines (dotted lines) for $H_{irr}^{ \perp}(T)$ (circles) and $H_{irr}^{\parallel} (T)$ data (circles), respectively, with the parameters $c_f^{\perp}=4.3$,  $q^{\perp}_0=5.0$,  $c_f^{\parallel}=3.4$ and $q^{\parallel}_0=2.0$. 
%
%
%
%
The discrepancy between the data and the fits led us to review the assumption of Blatter and Ivlev model \cite {Blatter}, concerning the equal weights of the quantum and thermal effects. As can be expected from the change of the magnetic structure along the irreversibility line (parameters in $q$ for reality), affected by the crystallographic structure, defects, inhomogeneities, and compositional deviations, we take a temperature dependent weight function $w(t)$, representing the relative importance of the thermal and quantum effects. 
Introducing this weight function, we use the following expression (for both orientations) for an effective quantum parameter $q(t)$:
%
%
\begin{equation}
q(t)=w(t)q_0(1-t)^{1/2}=q_{0}(1-t)^p,
\label{eq5}
\end{equation}
%
%
where $w$ is the weight function with $w(0)=1$ (no thermal effects at $T=0$) and $w(1)=0$ (no quantum effects above $T_c$), and $p$ an adjustable free parameter.  

The solid lines in Fig.~\ref{f3}(a) and ~\ref{f3}(b) represent the corresponding fits , in good agreement with the experimental data, with $c_f^{\perp}=4.3$,  $q^{\perp}_{0}=5.0$,  $c_f^{\parallel}=3.4$ and $q^{\parallel}_{0}=2.0$ the same values as before.  
The solid lines in the insets of Fig.~\ref{f3}(a) and ~\ref{f3}(b) represent the corresponding temperature dependence of the effective quantum parameters together with the corresponding $p$ values. Note that $p=1/2$ corresponds to equal weights [$w(t)\equiv 1$] of the thermal and quantum effects.  The differences between the curves with and without the temperature dependence $w(t)$ suggest that the quantum effects are reduced with a large $p$ value when $0< t < 1$, but still contribute substantially to the vortex motion and affect the irreversibility line.  
From the fitting, we determine $\mu _{0} H_{irr}^{ \perp}  
( {0} ) = 8.74$ T and $\mu _{0} H_{irr}^{\parallel} ( {0} ) 
= 17.3$ T, values similar to our previous estimations. With $c_{L}^{ 
\perp}  = 0.25$, we have $G^\perp \approx 5.3\times 10^{-4}$. As $G^{\parallel}/G^{ \perp}  \approx \gamma ^{2}$, we determine $G^{\parallel} \approx 1.2 \times 10^{ - 3}$ and $c_{L}^{\parallel} \approx 0.27$. One will notice that changing the $c_{L}^{ \perp}$ value will affect the values of $c_{L}^{\parallel} $, $G^{ \perp} $, and $G^{\parallel}$; the lower the $c_{L}^{ \perp}  $ value, the lower the $c^\parallel_L$, $G^{ \perp} $, and $G^{\parallel}$ values. 

%
%
\begin{figure}
\includegraphics
[width=0.88\columnwidth]
{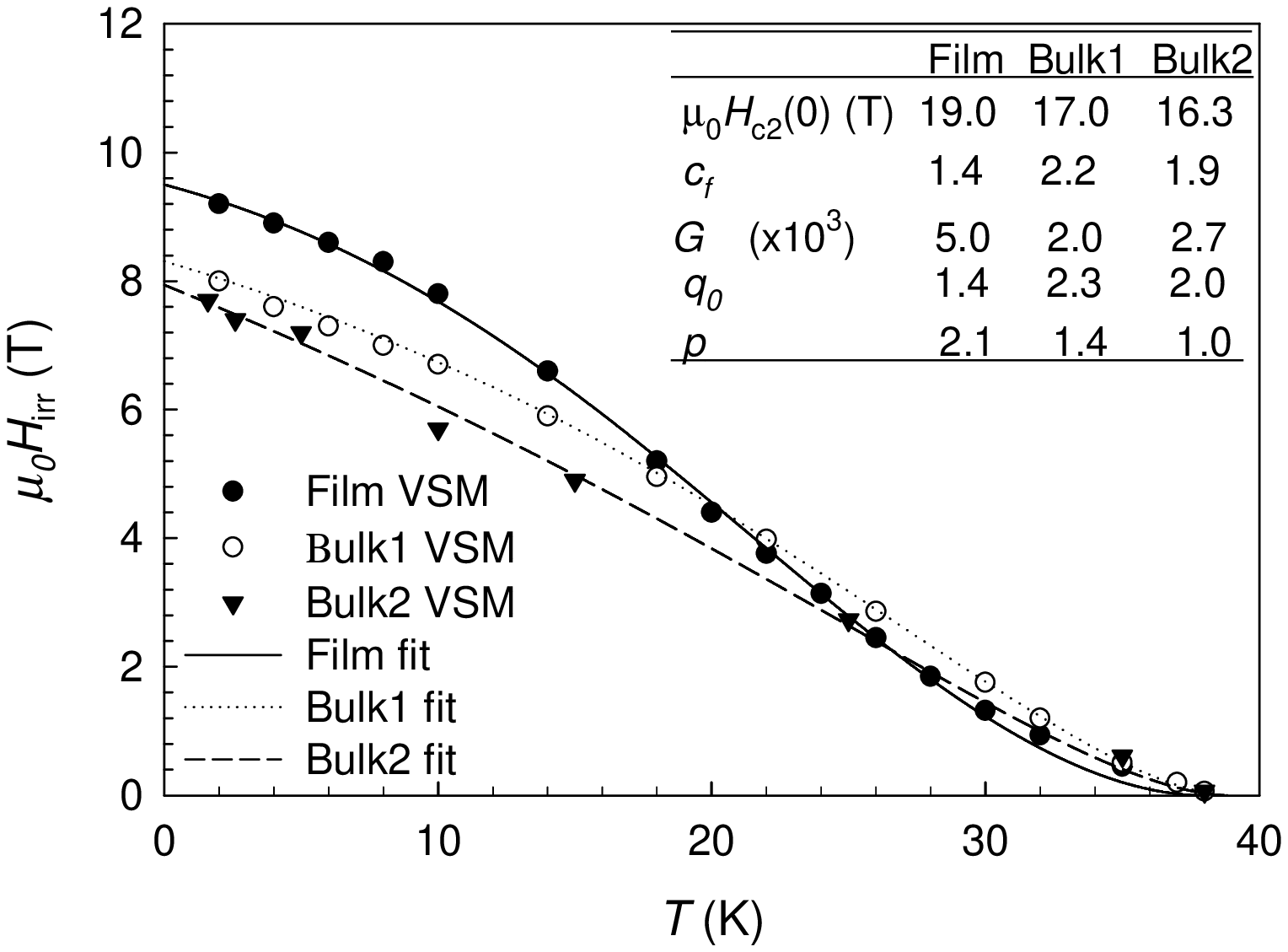}\caption{\label{f4}  Theoretical fits of $H_{irr}(T)$ data of several MgB$_2$ superconductors (one thin film and two bulk materials) \cite {Wen}. The table gives the relative parameters of the samples in corresponding fits, for which we simply assume $H_{c2}(0)\sim 2 H_{irr}(0)$, and $c_L=0.25$.}
\end{figure} 
%
%
Figure~\ref{f4} represents fits of irreversibility lines of several other MgB$_2$ samples [one thin film ($T_c \approx 38$ K) and two MgB$_2$ bulk materials ($T_c \approx 39$ K)] using a vibrating sample magnetometer (VSM) \cite {Wen}, in good agreement with the experimental data. The inset table presents the corresponding fitting parameters.
The following differences have to be considered when comparing the data as their crystallographic structures and defects are rather different. Both thin films, having relatively large $p$ values when comparing with bulk samples, are mainly $c$-axis oriented with textured structures. The bulk samples are non-oriented polycrystals with $T_c \approx $ 39 K,  a typical  value for MgB$_2$. All the VSM tested samples with relatively high $T_c$s show relatively small $q_0$ and $c_f$ values (large $G$ values). 

Based on the study of quantum effects at low temperatures in HTSCs, Ivlev, Ovchinnikov, and Thompson proposed a crossover temperature $T_0$ ($0<T_0<T_c$) to separate thermal and quantum regimes \cite {Ivlev,Smith}, where $T_0$ can be much smaller than $T_c$ in HTSCs \cite {Zhang,Yeshurun,Nowak,Monier,Hoekstra,Stein}. Our study, however, suggests that in MgB$_2$ superconductor $T_0\to T_c$. This difference between HTSCs and MgB$_2$ implies that the form of $w(t)$ in the study is not universal for all superconductors, but is superconductor dependent, and thus may account for the $H_{irr}(T)$ curvature difference between HTSCs and MgB$_2$.

To summarize, a $c$-axis oriented MgB$_{2}$ thin film is studied in perpendicular and parallel fields up to 23 T yielding resistive $H_{irr}$ and 
$H_{c2}$ in both directions. The anisotropy factor is determined to 
be around 1.5 by extrapolating the $H_{c2}^{\parallel} ( {T} )$ 
and $H_{c2}^{\perp}$ curves to 
$T = 0$. At low temperatures, $H_{i}^{ \perp}  ( {T} )$ and 
$H_{irr}^{\parallel}(T)$ are nearly temperature independent, and 
large differences between the values for $H_{irr}(T)$ and 
$H_{c2}(T)$ are observed. The analysis concludes that the critical field data are related to quantum fluctuations of the vortex matter in this new superconductor. We propose the effective quantum parameter $q(t)$ under the framework of quantum fluctuations. The relative contributions of quantum and thermal effects are evaluated for the irreversibility line. Due to the dependence of the vortex parameters on magnetic field and temperature, the relative weights of the two contributions are temperature dependent. Complementary studies of $H_{irr}(T)$ data of other MgB$_2$ samples \cite {Wen} give a strong support to the analysis.

This work has been financially supported by the National Science Foundation 
of China, PAI 4/10 (Belgium), Communaute Francaise de Belgique, the Ministry 
of Science and Technology of China, and National Science Foundation of USA.

\end{document}